\documentclass[11pt,a4paper]{article}
\usepackage{xcolor}
\usepackage{fullpage}
\usepackage{amsmath,amsthm,amssymb}
\usepackage{epsfig}
\usepackage{pst-grad} 
\usepackage{pst-plot} 

\newtheorem{theorem}{Theorem}

\newtheorem{lemma}[theorem]{Lemma}
\long\def\symbolfootnote[#1]#2{\begingroup
\def\thefootnote{\fnsymbol{footnote}}\footnote[#1]{#2}\endgroup}

\def\sci{{\chi_s'}}

\def\calC{{\mathcal{C}}}
\newcommand*\samethanks[1][\value{footnote}]{\footnotemark[#1]}
\author{Valentin Borozan\thanks{{L.R.I.,  B\^at. 650, Universit\'e Paris-Sud, Orsay, France}}~\thanks{valentin.borozan@gmail.com}\and Leandro Montero\samethanks[1]~\thanks{lmontero@lri.fr} \and Narayanan N\thanks{Indian Institute of Technology Madras, India.}~\samethanks[1]~\thanks{naru@iitm.ac.in}}
\begin{document}
\title{\bf Further results on strong edge-colourings in outerplanar graphs}
\date{} 
\maketitle

\begin{abstract} An edge-colouring is {\em strong} if every colour class is an
induced matching. In this work we give a formulae that determines either the optimal or the optimal plus one strong chromatic index of bipartite
outerplanar graphs. Further, we give an improved upper bound for any 
outerplanar graph which is close to optimal. All our proofs yield efficient
algorithms to construct such colourings.  \bigskip 


\end{abstract}

\section{Introduction}   

Given a simple undirected graph $G$, let $V(G)$ and $E(G)$ respectively denote
the vertex set and the edge set of $G$. A {\it proper  $k$-edge-colouring} of
$G$ is a map $\calC:E(G) \mapsto [k]$ such  that adjacent edges (edges of $G$
sharing a common vertex) receive different colours (numbers), where  $[k] =
\{1, 2, \ldots, k\}$. The smallest positive integer $k$ such that  $G$ admits a
proper  $k$-edge-colouring is known as the {\em chromatic index} of $G$ and is
denoted $\chi'(G)$. 

An {\em induced matching} $M$ in $G$ is a matching such that $G[V(M)]=M$. That
is, the subgraph of $G$ induced by the vertices of $M$ is $M$ itself.  A proper
edge-colouring is a {\em strong edge-colouring} if every colour class is an
induced matching in $G$. In other words, for any edge $e=uv$, the sets of
colours {\em seen} by $u$ and $v$ have exactly one colour in common (in an edge-colouring 
we say that a vertex {\em sees} colour $c$ if $c$ is assigned to
any of the edges incident to it). That is, the distance between any two
edges having the same colour is at least two.  The minimum positive integer $k$
such that $G$ admits a strong  $k$-edge-colouring is called the {\em strong
chromatic index} of $G$ denoted $\sci(G)$.
The degree of a vertex $v$ is denoted $d(v)$. 
An edge incident to a vertex of degree one is called a {\em pendant edge}. Let $\Delta=\Delta(G)$ denote the maximum degree of vertices of $G$. 

A graph $G$ is {\em outerplanar} if it has a planar embedding in which all vertices are incident to the infinite face.
We define a {\em puffer graph} as a graph obtained by
adding some (possibly empty) pendant edges to each vertex of a cycle or
adding a common neighbour to two consecutive vertices of the cycle. Notice
that for an outerplanar graph, at most one such vertex can be added.
All graphs are assumed to be connected.

The strong edge-colouring has a long history and has lead to many well known 
conjectures. Some of the many unsolved conjectures include $\sci(G)\le
5\Delta^2/4$ for all graphs,  $\sci(G) \le \Delta^2$ for bipartite graphs and
$\sci(G) \le 9$ for $3$-regular planar graphs. See the open problems pages of
Douglas West~\cite{westopen} for more details.

Molloy and Reed~\cite{mrstrong} proved a conjecture by Erd\H{o}s and
Ne\v{s}et\v{r}il {(see~\cite{sci-tuza}) } that for large $\Delta$, there is a
positive constant $c$ such that $\sci(G) \le (2-c)\Delta^2$.
Mahdian~\cite{mahdian} proved that for a $C_4$-free graph $G$, $\sci(G) \le
(2+o(1))\Delta^2/\ln \Delta$.

For integers $0\le\ell\le k \le m$, $S_m(k,\ell)$ is the bipartite graph with
vertex set $\{x \subseteq [m]\colon |x|=k$ or $\ell\}$ and a $k$-subset $x$ is
adjacent to an $\ell$-subset $y$  if $y \subseteq x$.  Quinn and
Benjamin~\cite{quinn} proved that $S_m(k,\ell)$ has strong chromatic index $m
\choose k-\ell$.  The {\it $\Theta$-graph} $\Theta(G)$ of a partial cube $G$ 
(distance-invariant subgraph of some $n$-cube), is the intersection graph of
the equivalence classes of the {\it Djokovi\'c-Winkler relation} $\Theta$
defined on the edges of $G$ such that $xy$ and $uv$ are in relation $\Theta$ if
$d(x,u)+d(y,v) \ne d(x,v)+d(y,u)$.  \v{S}umenjak~\cite{tadeja} showed that the
strong chromatic index of a tree-like partial cube graph $G$ is at most the
chromatic number of $\Theta(G)$.

Faudree, Gy\'arf\'as, Schelp and Tuza~\cite{tuzaim} proved that for graphs
where all cycle lengths are multiples of four, $\sci(G)\le\Delta^2$. They
mention that this result probably could be improved to a linear function of the
maximum degree. Brualdi and Quinn~\cite{incidstrong} improved the upper bound
to $\sci(G)\le\alpha\beta$ for such graphs, where $\alpha$ and $\beta$ are the
maximum degrees of the respective partitions. Nakprasit~\cite{scibp} proved
that if $G$ is bipartite and the maximum degree of one partite set is at most
$2$, then $\sci(G)\le2\Delta$.

In this work we improve the upper bound of $3\Delta-3$ for general outerplanar graphs given in a recent work~\cite{outersec}.
We also obtain either the exact value of $\sci(G)$ or the exact value plus one for bipartite outerplanar graphs. 
The following is our main result.

\begin{theorem}\label{thm:oup} Let $G$ be an outerplanar graph. Then
$\sci(G)=\max\{\max_{uv\in E} d(u)+d(v)-1, \max_{H\in \cal{P}} \sci(H)\}$, where $\cal{P}$ is the set of all
puffer subgraphs of $G$. Moreover, if $G$ is bipartite then $\sci(G)$ is either $\max_{uv\in E} d(u)+d(v)-1$ or at most $\max_{uv\in E} d(u)+d(v)$.
\end{theorem}

We also give efficient algorithms to produce strong edge-colourings of such
classes of graphs satisfying the above bounds.

\section{Outerplanar graphs}

A {\em block} is a maximal connected component
without a cut-vertex. A {\em block decomposition} of a graph $G$ is a
partition of $G$ into its blocks. Notice that each component is either a
maximally $2$-connected subgraph or a single edge. We define an {\em end block}
of a graph as a {\em $2$-connected} block that contains a unique cut-vertex
which separates it from all the other $2$-connected blocks (if exists). Notice
that this definition of end block differs from the standard notion in order
to address some specific issues. 
 
An {\em ear} in $G$ to a subgraph $H$ is a simple path $P$ on at least three
vertices with end-points in $H$ such that (1) none of the internal vertices of
$P$ are contained in $H$ and (2) $P$ along with the segment between its end-points in $H$ forms an induced cycle.  An
{\em ear decomposition} of a $2$-connected subgraph is a partition of its edges
into a sequence of ears where the first ear is an induced cycle. It is easily
seen that for a $2$-connected outerplanar graph, there is an ear decomposition
where each ear contains at least one internal vertex and the endpoints of every
ear are adjacent in the preceding graph (if not, the outerplanarity property is
affected). Further notice that when the graph is bipartite outerplanar any
added ear has an even number of internal vertices. Any such ear (which forms an
induced cycle) together with the edges incident to it forms a puffer graph we
defined earlier. Thus, we first show an upper bound for the puffer graphs.

\subsection{Puffer graphs}

Note that to compute the strong chromatic index we suppose that the puffer
graph only has pendant edges (no common neighbours forming a triangle) since we
can always split any common neighbour of adjacent vertices of the cycle to two
pendant edges which does not affect the colouring. 

The following lemma gives bounds for the puffer graphs.

\begin{lemma}\label{lem:pg} Let $G$ be a puffer graph and $C$ its cycle. We have the following according to the cycle length $|C|$.

\begin{enumerate} 
\item $\sci(G) =  d(u)+d(v)+d(w)-3$ if $|C|=3$, ${u,v,w\in C}$.
\item $\sci(G) = \max_{uv \in E(C)} d(u)+d(v)$ if $|C|=4$.
\item $\sci(G) = 5$ if $G=C_5$.
\item $\sci(G) = \max_{u \in C} d(u)+2$ if $|C|=5$ and either only a single vertex or exactly two vertices at distance $2$ have pendant edges.
\item $\sci(G) = \max_{uv\in E(C)} d(u)+d(v)-1$  if $|C|=5$, at least one vertex has at most $1$ pendant edge and not in cases $3)$ and $4)$.
\item If $|C|=5$ and every vertex has at least $2$ pendant edges. Let $u,v$ be the vertices where the $\max_{u_1u_2 \in E(C)}d(u_1)+d(u_2)$ is reached and 
let $x,y$ and $z$ be the rest of the vertices. Call $\eta = \lceil{d(x)+d(y)+d(z)-d(u)-d(v)-3\over{2}}\rceil$. Then
\begin{equation*}
  \sci(G) \le 
      \begin{cases}
	d(u)+d(v)-1 \text{ if } d(u)+d(v) \geq d(x)+d(y)+d(z)-3 \\
	d(u)+d(v)-1 + \eta \text{ otherwise}
    \end{cases}
\end{equation*}
\item If $G=C_k$, $k \geq 6$, then
\begin{equation*}
  \sci(G) = 
      \begin{cases}
	3 \text{ if } k\equiv 0 (\bmod 3)\\
	4 \text{ otherwise }
    \end{cases}
\end{equation*}

\item Let $G\neq C_{2k}$, $k\geq 3$. Let $C_{2k}=v_1v_2\ldots v_{2k}v_1$. Let $u,v$ be the vertices where the $\max_{u_1u_2 \in E(C_{2k})}d(u_1)+d(u_2)$ is 
reached and suppose without losing generality that $u=v_1$, $v=v_2$ and at least $u$ has a non-empty set of pendant edges.
\begin{enumerate}
 \item If $2k \equiv 0 (\bmod 3)$ then $\sci(G) = d(u)+d(v)-1$. 
 \item If $2k \equiv 2 (\bmod 3)$ then 
\begin{itemize}
\item $\sci(G) \leq d(u)+d(v)$ if there exists another pair of vertices $v_jv_{j+1} \in E(C_{2k})$ such that $d(u)+d(v) = d(v_j)+d(v_{j+1})$, 
$j$ is even and $d(v_{j+1})=2$ (clearly this implies that $d(v_{j-1})=2$).
\item $\sci(G) = d(u)+d(v)-1$ otherwise.
\end{itemize}
\item If $2k \equiv 1 (\bmod 3)$ then
\begin{itemize}
\item $\sci(G) \leq d(u)+d(v)$ if for every vertex $w \in C_{2k}$ we have $d(w) \leq 3$.
\item $\sci(G) \leq d(u)+d(v)$ if there exists another pair of vertices $v_jv_{j+1} \in E(C_{2k})$ such that $d(u)+d(v) = d(v_j)+d(v_{j+1})$, 
$j$ is even and $d(v_{j+1})\leq 3$ (clearly this implies that $d(v_{j-1}) \leq 3$).
\item $\sci(G) = d(u)+d(v)-1$ otherwise.
\end{itemize}
\end{enumerate}

\item Let $G\neq C_{2k-1}$, $k\geq 4$. Let $C_{2k-1}=v_1v_2\ldots v_{2k-1}v_1$.
Let $u,v$ be the vertices where the $\max_{u_1u_2 \in E(C_{2k-1})}d(u_1)+d(u_2)$
is reached and let $x,y$ and $z$ be the consecutive vertices of $C_{2k-1}$ not
considering $u$ and $v$ where the $\min_{s_1,s_2,s_3 \in C_{2k-1}}d(s_1)+d(s_2)+d(s_3)$ is attained. 
Suppose without losing generality that $v_1=x$, $v_2=y$ and $v_3=z$. Let $\eta = \lceil{d(x)+d(y)+d(z)-d(u)-d(v)-2\over{2}}\rceil$. 

\begin{enumerate}
 \item If $2k-1 \equiv 0 (\bmod 3)$ then
\begin{equation*}
      \begin{cases}
	\sci(G) = d(u)+d(v)-1 \text{ if } d(u)+d(v) \geq d(x)+d(y)+d(z)-2 \\
	\sci(G) \leq d(u)+d(v)-1 + \eta \text{ otherwise }
    \end{cases}
\end{equation*}
\item If $2k-1 \equiv 2 (\bmod 3)$ then
\begin{itemize}
\item $\sci(G) \leq d(u)+d(v)$ if $d(u)+d(v) \geq d(x)+d(y)+d(z)-2$ and there exists another pair of vertices $v_jv_{j+1} \in E(C_{2k-1})$ such that 
$d(u)+d(v) = d(v_j)+d(v_{j+1})$, $j$ is odd and $d(v_{j+1})=2$ (clearly this implies that $d(v_{j-1})=2$).
\item $\sci(G) = d(u)+d(v)-1$ if $d(u)+d(v) \geq d(x)+d(y)+d(z)-2$ and we are not in the previous case.
\item $\sci(G) \leq d(u)+d(v)-1 + \eta$ if $d(u)+d(v) < d(x)+d(y)+d(z)-2$.
\end{itemize}
\item If $2k-1 \equiv 1 (\bmod 3)$ then 
\begin{itemize}
\item $\sci(G) \leq d(u)+d(v)$ if for every vertex $w \in C_{2k-1}$ we have $d(w) \leq 3$.
\item $\sci(G) \leq d(u)+d(v)$ if $d(u)+d(v) \geq d(x)+d(y)+d(z)-2$ and there exists another pair of vertices $v_jv_{j+1} \in E(C_{2k-1})$ such that 
$d(u)+d(v) = d(v_j)+d(v_{j+1})$, $j$ is odd and $d(v_{j+1})\leq 3$ (clearly this implies that $d(v_{j-1}) \leq 3$).
\item $\sci(G) = d(u)+d(v)-1$ if $d(u)+d(v) \geq d(x)+d(y)+d(z)-2$ and we are not in the previous case.
\item $\sci(G) \leq d(u)+d(v) + \eta$ if $d(u)+d(v) < d(x)+d(y)+d(z)-2$ and and there exists another pair of vertices $v_jv_{j+1} \in E(C_{2k-1})$ such that 
$d(u)+d(v) = d(v_j)+d(v_{j+1})$, $j$ is odd and $d(v_{j+1})= 3$ (clearly this implies that $d(v_{j-1}) \leq 3$).
\item $\sci(G) \leq d(u)+d(v)-1 + \eta$ if $d(u)+d(v) < d(x)+d(y)+d(z)-2$ and we are not in the previous case.
\end{itemize}
And same bounds plus $1$ if $|C_{2k-1}|=7$.
\end{enumerate}

\end{enumerate}

\end{lemma}

We give below a proof of the above lemma. We then show how to colour any
outerplanar graph in a strong manner using the lemma and the ear decomposition.
We start by colouring the first ear (a cycle) with its incident edges (which
forms a puffer graph) and extend that colouring to the next puffer graph (next
ear with its incident edges). 

\noindent{\bf Proof of Lemma~\ref{lem:pg}.} 

The proof is trivial for statements $1)$ through $4)$. Statement $5)$ can also be easily verified.  

For $6)$ we note that every vertex of the cycle has at least $2$ pendant edges.
Let $uvxyz$ be the vertices of the $C_5$ in a cyclic order. We colour the cycle
with colours $1$ to $5$ starting at the edge $uv$. Then we colour one
uncoloured incident edge of each  vertex with the only possible colour among
the used ones (keeping the strong colouring property). Thus we have
$d(u)+d(v)-6$ uncoloured edges incident to $u$ and $v$, and $d(x)+d(y)+d(z)-9$
uncoloured edges incident to $x,y$ and $z$. Suppose that $d(u)+d(v)-6 \geq
d(x)+d(y)+d(z)-9$, i.e., $d(u)+d(v) \geq d(x)+d(y)+d(z)-3$.  We use $d(u)-3$
new colours to colour the uncoloured edges at $u$ and $d(v)-3$ new colours for
the ones at $v$. We remark that this is the only possibility to keep the strong
colouring property. Clearly $d(x) \leq d(u)$ and $d(z) \leq d(v)$. We colour
the uncoloured  edges at $x$ and $z$ from the set of colours used
at $u$ and $v$ respectively.  Since  $d(x)+d(y)+d(z)-3 \le d(u)+d(v)$, we
notice that there are enough colours left to colour the edges incident to $y$.
Since we use only $d(u)+d(v)-1$ colours, the bound is optimal in this case. 

Now suppose that $d(u)+d(v) < d(x)+d(y)+d(z)-3$. As before, we colour the
$d(u)-3$ edges at $u$ and the $d(v)-3$ edges at $v$ with new colours. Now we
introduce an additional $\eta$ new colours and colour as many edges incident
to both $x$ and $z$ (we can verify that both $x$ and $z$ have at least $\eta$
uncoloured edges in this case).  Then for the remaining edges at $x$ we use at
most $d(x)-3-\eta$ colours used at $u$ and for the ones at $z$ use at most
$d(z)-3-\eta$  colours used at $v$.  As before it is not difficult to see that
we have enough colours left to colour the edges incident to $y$.

For statement $7)$, we colour the cycle in the following way. If $k\equiv 0
(\bmod 3)$ we use colours $1,2$ and $3$ repeatedly for the cycle and we are done.
If $k\equiv 1 (\bmod 3)$, we colour one edge with colour $4$ and then repeatedly
with colours $1,2$ and $3$. Finally, if $k\equiv 2 (\bmod 3)$, we colour the
first $5$ edges with colours $4,1,2,3,4$ and then repeatedly with colours
$1,2$ and $3$. Again, it works since $k>7$.

For $8a)$, colour the edges of the cycle repeatedly with colours $1,2$ and $3$ starting from the edge $v_1v_2$. 
Clearly the cycle is strong edge-coloured since $2k \equiv 0 (\bmod 3)$. 
Now introduce a set of new colours $A$, where $|A|=d(v_1)-2$ and for each odd vertex on the cycle
colour its uncoloured incident edges with colours from $A$ using the least
permissible colour. Then do the same for each even vertex on the cycle using
another set of new colours $B$, where $|B|=d(v_2)-2$. If there are not more uncoloured edges we are done.  

Suppose now that there exists one vertex $v_j$, $j$ odd ($j$ even is similar) such that $d(v_j) > d(v_1)$.
Therefore there are $d(v_j) - d(v_1)$ edges to colour incident to $v_j$. 
We know that $d(v_1)+d(v_2) \geq d(v_j)+d(v_{j+1})$ and $d(v_1)+d(v_2) \geq d(v_j)+d(v_{j-1})$.
So if we suppose (without losing generality) that $d(v_{j+1})\geq d(v_{j-1})$ then $d(v_2)-d(v_{j+1}) \geq d(v_j)-d(v_1)$.
Therefore we have $d(v_2)-d(v_{j+1})$ colours from $B$ not used neither at $v_{j-1}$ nor at $v_{j+1}$ and then we can 
use them to colour the remaining $d(v_j) - d(v_1)$ edges at $v_j$.
Clearly this edge-colouring is strong and we used $3+|A|+|B|=3+d(u)-2+d(v)-2=d(u)+d(v)-1$ colours, which is
optimal since $d(u)+d(v)-1$ is also a lower bound.

For $8b)$, we colour the cycle with colours $1,2,3$ starting at the edge $v_1v_2$ until the 
edge $v_{2k-4}v_{2k-3}$ and for the four remaining edges we use colours (respecting the cycle ordering) $4,3,2,4$. Then, 
by the way we coloured the cycle we have that for each odd vertex $v_i$ in the cycle there is one available colour among the colours 
$\{1,2,3,4\}$ to use at its uncoloured incident edges. We proceed to colour this edges with that colour. 
As in $8a)$, introduce a set of new colours $A$, but now $|A|=d(v_1)-3$ and for each odd vertex on the cycle
colour its uncoloured incident edges with colours from $A$ using the least
permissible colour. For each even vertex on the cycle do the same using 
another set of new colours $B$, where $|B|=d(v_2)-2$. If there are no more edges to colour we are done.

Suppose now that there exists one vertex $v_j$ with uncoloured edges incident to it. Suppose also that $d(v_1)+d(v_2) = d(v_j)+d(v_{j+1})$, 
$j$ is even and $d(v_{j+1})=2$. Therefore $d(v_j) > d(v_2)$. Now, since $d(v_{j+1})=2$ (and also $d(v_{j-1})=2$). We can use the $d(v_1)-3$ colours
from $A$ for the remaining uncoloured edges at $v_j$. By this we colour $2 + d(v_1)-3+d(v_2)-2 = d(v_1) + d(v_2) -3$ edges at $v_j$.
However, since $d(v_1)+d(v_2) = d(v_j)+d(v_{j+1})$ we have that $d(v_j) = d(v_1)+d(v_2) - d(v_{j+1}) = d(v_1)+d(v_2) - 2$ therefore we need one new colour more 
to finish colouring $v_j$ in a strong manner since $v_j$ sees the four colours used at the cycle ($j$ is even). By this we used 
$4 + |A| + |B| + 1 = d(u)+d(v)$ colours and then $\sci(G) \leq d(u)+d(v)$\footnote{We remark that if there are few vertices $v_j$ with that property,  
we can maybe recolour the cycle with four colours such that each $v_j$ sees only three colours among the four used at the cycle. Then this new colour 
would not be necessary and we would have $\sci(G) = d(u)+d(v)-1$. Nevertheless, there are graphs where we cannot do this and therefore 
$\sci(G) = d(u)+d(v)$.}.

To finish this case if we are not in the previous conditions for $v_j$ then at least one condition among these three is true: 
(1) $d(u)+d(v) > d(v_j)+d(v_{j+1})$, (2) $j$ is odd or (3) $d(v_{j+1})\geq 3$. 
In each of these cases the way to colour the remaining edges at $v_j$ is similar to the case $8a)$ and we obtain $\sci(G) = d(u)+d(v)-1$ which is optimal.

For $8c)$, the case where every vertex $w \in C_{2k}$ has $d(w) \leq 3$ can be easily verified.
Otherwise we colour the cycle repeatedly with colours $1,2,3$ starting at the edge $v_1v_2$ until the 
edge $v_{2k-6}v_{2k-5}$ and for the six remaining edges we use colours (respecting the cycle ordering) $5,3,4,5,2,4$. Then, 
by the way we coloured the cycle we have now that for each odd vertex $v_i$ in the cycle there are two available colours among the colours 
$\{1,2,3,4,5\}$ to use at its uncoloured incident edges. We colour them with those colours. 
Similar to $8b)$, for each odd vertex on the cycle colour its uncoloured incident edges with a new set of colours $A$ where $|A|=d(v_1)-4$ and 
for each even vertex on the cycle do the same using another set of new colours $B$ where $|B|=d(v_2)-2$. If there are not more uncoloured edges we are done. 

Suppose now that there exists one vertex $v_j$ with uncoloured edges incident to it. Suppose also that $d(u)+d(v) = d(v_j)+d(v_{j+1})$, 
$j$ is even and $d(v_{j+1})\leq 3$. Therefore $d(v_j) > d(v_2)$. Suppose first that $d(v_{j+1})=2$ (and also $d(v_{j-1})=2$). 
We can use the $d(v_1)-4$ colours from $A$ for the remaining uncoloured edges at $v_j$. By this we colour $2 + d(v_1)-4+d(v_2)-2 = d(v_1) + d(v_2) -4$ 
edges at $v_j$. However, since $d(v_1)+d(v_2) = d(v_j)+d(v_{j+1})$ we have that $d(v_j) = d(v_1)+d(v_2) - d(v_{j+1}) = d(v_1)+d(v_2)-2$ therefore we 
still need to colour two edges. Now, since $v_j$ sees four colours among the five used at the cycle and $d(v_{j+1})=2$, $d(v_{j-1})=2$, 
we colour one of its two remaining edges with that colour. For the last one, we need to use a new colour and we finish colouring $v_j$.
Suppose last that $d(v_{j+1})=3$ (and then $d(v_{j-1})\leq 3$). As before, if we use the colours from $A$ for the uncoloured edges at $v_j$, 
we colour $d(v_1) + d(v_2) -4$ edges at $v_j$ but now $d(v_j) = d(v_1)+d(v_2)-3$. Then we still have to colour one more edge at $v_j$. 
For this one, we need to use a new colour since the colour that $v_j$ does not see among the five used at the cycle is used either at $v_{j+1}$ or
at $v_{j-1}$ or at both. 
In both cases this we used $5 + |A| + |B| + 1 = d(u)+d(v)$ colours and then $\sci(G) \leq d(u)+d(v)$\footnote{Same remark as case $8b)$ applies.}.

If there is no $v_j$ satisfying these three conditions, we colour its remaining edges as in cases $8a)$, $8b)$ and we obtain $\sci(G) = d(u)+d(v)-1$.

For $9a)$, colour the edges on the cycle from the edge $v_1v_2$ with colours $1$,$2$ and $3$ repeatedly
Clearly, the cycle is strong edge-coloured 
since $2k-1 \equiv 0 (\bmod 3)$. We will colour the rest of the graph as in $8a)$ but considering the vertices $v_1$ and $v_2$ as a single vertex (say $v_2$).
Observe that there are $d(u)+d(v)-4$ uncoloured edges at $u$ and $v$ and $d(x)+d(y)+d(z)-6$ uncoloured ones at $x,y$ and $z$.
Suppose first then that $d(u)+d(v)-4 \geq d(x)+d(y)+d(z)-6$, that is, $d(u)+d(v) \geq d(x)+d(y)+d(z)-2$. 
Then, introduce a set of new colours $A$ where $|A|=d(u)-2$ and another set of new colours $B$ where $|B|=d(v)-2$. 
Now colour the rest of the edges as in $8a)$ considering $v_1$ and $v_2$ as a single vertex. Clearly, this
leads to a strong edge-colouring of $G$ following same arguments as in $6)$ and
$8a)$. We use $d(u)+d(v)-1$ colours.
Second, suppose that $d(u)+d(v)<d(x)+d(y)+d(z)-2$. We use $\eta$ (as defined
earlier) new colours to colour a subset of $\eta$ edges incident to each of
$v_1$ and $v_3$.  Again it is easily seen that $d(v_1)$ and $d(v_3)$ are
at least $\eta$ using the assumed inequalities.  Finally to colour the rest of
the edges we proceed as in the first case. As before, the colouring is strong
by $6)$ and $8a)$. We use $d(u)+d(v)-1+\eta$ colours as desired. 

For $9b)$, we colour the cycle repeatedly with colours $1,2,3$ starting at the edge $v_1v_2$ until the 
edge $v_{2k-7}v_{2k-6}$ and for the six remaining edges we use colours (respecting the cycle ordering) $4,1,3,4,2,3$. 
We can observe that for each even vertex $v_i$ in the cycle there is one available colour 
among the colours $\{1,2,3,4\}$ to use at its uncoloured incident edges. Then combining cases $8b)$ and $9a)$ the result holds.
We remark that the case that $d(u)+d(v) < d(x)+d(y)+d(z)-2$ and there exists another pair of vertices $v_jv_{j+1} \in E(C)$ such that 
$d(u)+d(v) = d(v_j)+d(v_{j+1})$, $j$ is odd and $d(v_{j+1})=2$, is not possible since that would contradict the fact that $x,y$ and $z$ were 
the consecutive vertices that minimise the sum of degrees or that $u$ and $v$ maximised it.

For $9c)$, we have a similar situation as case $9b)$ but here we colour the cycle repeatedly with colours $1,2,3$ starting at the edge $v_1v_2$ until the 
edge $v_{2k-9}v_{2k-8}$ and for the seven remaining edges we use colours (respecting the cycle ordering) $5,1,4,5,3,4,2,3$
(for the special case $|C|=7$ we colour $5,1,4,5,3,4,2$). 
Now for each even vertex $v_i$ in the cycle there are two available colours 
among the colours $\{1,2,3,4,5\}$ to use at its uncoloured incident edges (for the special case $|C|=7$ we may need one more colour since 
by the lenght of the cycle there exists one even vertex without this property). 
Then we combine cases $8c)$ and $9b)$. We remark that as in the previous case, 
we cannot have at the same time $d(u)+d(v) < d(x)+d(y)+d(z)-2$ and the existence of another pair of vertices $v_jv_{j+1} \in E(C)$ such that 
$d(u)+d(v) = d(v_j)+d(v_{j+1})$, $j$ is odd and $d(v_{j+1})=2$. However, we can for $d(v_{j+1})=3$, therefore we might need one more colour.

\noindent{\bf Proof of Theorem~\ref{thm:oup}}

We observe that given a block decomposition of an outerplanar graph and then an ear decomposition of each block, every ear
together with the edges incident to it forms a puffer graph. Then, adding the ears in
the order of the decomposition, each new ear joins two adjacent vertices. Since only the edges incident
to two adjacent vertices of the new ear is precoloured, we note that we can
simply extend the colouring to the new puffer graph (as the precoloured edges
all get distinct colours in both cases). The upper bound for outerplanar graphs
is now clear by maximising over all puffer graphs and over all pairs of
adjacent vertices (the latter is a trivial lower bound). When the graph is
bipartite, this gives either the exact value or the exact value plus one colour. \hfill $\blacksquare$

The proof itself gives the algorithm to obtain such a colouring and it is easy
to see that it takes sub-quadratic time.

\section{Remarks}

In this work we consider outerplanar graphs. We give a formulae to find either the exact value or the exact value plus one of the strong
chromatic index for bipartite outerplanar graphs. We also improve the
upper bound for the general outerplanar graphs from the $3\Delta-3$
stated in~\cite{outersec}. 

A recent work~\cite{opalgo} gives an algorithm to find the strong chromatic
index of any maximal outerplanar graph. However, notice that when you extend the
graph to maximal outerplanar, the maximum degree and the index can increase.
We provide an algorithm to colour any outerplanar graph with a number of colours
close to the optimum and for bipartite outerplanar graphs with the optimum (or just one more) number of colours.

In some special cases of the general outerplanar graph (where we use $\eta$
extra colours) we are not able to show the optimality of the bounds. We believe
that it is very close to the exact bound within an additive factor of a small
constant. It would be interesting to prove if our bounds are optimal and if
not, to find a way to close the gap.

\bibliographystyle{abbrv} \bibliography{biblio}

\end{document}